\documentclass{ws-ijmpa}

\begin{document}

\markboth{Kim Maltman}{QCD Sum Rule Constraints on $a_\mu$}

\catchline{}{}{}{}{}

\title{Resolving the $\tau$ vs. electroproduction discrepancy for the
$I=1$ vector spectral function and implications for the SM prediction
for $a_\mu$}

\author{Kim Maltman}
\address{Dept. Math and Stats, York Univ., 4700 Keele St.,
Toronto, ON CANADA M3J 1P3, and
CSSM, Univ. of Adelaide, Adelaide, SA AUSTRALIA 5005,
e-mail: kmaltman@yorku.ca}

%%%%%%%%%%%%%%%%%%%%%%%%%%%%%%%%%%%%%%%%%%%%%%%%%%%%%%%%%%%%
% You may repeat \author \address as often as necessary    %
%%%%%%%%%%%%%%%%%%%%%%%%%%%%%%%%%%%%%%%%%%%%%%%%%%%%%%%%%%%%

\maketitle

%\begin{history}
%\received{DAY MONTH YEAR}
%\revised{DAY MONTH YEAR}
%\end{history}

\begin{abstract}
Using only independent high-scale OPE input, we investigate
QCD sum rule constraints on two currently incompatible versions 
of the isovector vector spectral function, one obtained from 
electroproduction (EM) data, the other from hadronic $\tau$
decay data. Sum rules involving weighted integrals over the
spectral function, from threshold to a variable upper endpoint
$s_0$, are employed. It is shown that both the normalization and
slope with respect to $s_0$ of the EM spectral integrals 
disagree with the corresponding OPE expectations, while
both normalization and slope are in good agreement 
when hadronic $\tau$ decay data is used instead. These 
results favor determinations of the leading
hadronic vacuum polarization contribution to $a_\mu$
obtained using the $\tau$ decay data, and hence Standard
Model predictions for $a_\mu$ compatible with the
current experimental determination.
\end{abstract}


%%%%%%%%%%%%%%%%%%%%%%%%%%%%%%%%%%%%%%%%%%%%%%%%%%%%%%%%%%%%
% The main text of your paper	begins here	           %
%%%%%%%%%%%%%%%%%%%%%%%%%%%%%%%%%%%%%%%%%%%%%%%%%%%%%%%%%%%%

\section{Introduction}
The recent high-precision experimental determination of
$a_\mu =(g-2)_\mu$~\cite{bnlamu}
has led to intense interest in the corresponding Standard Model (SM)
prediction~\cite{review04}. 
The largest of the non-pure-QED SM contributions, that due to
the leading order hadronic vacuum polarization,
$[ a_\mu ]_{had}^{LO}$, can be obtained as an 
appropriately-weighted integral over the electromagnetic (EM) 
spectral function, $\rho_{EM}(s)$~\cite{amuintegral}. 
The $I=1$ component of $\rho_{EM}(s)$ is
related to that of its charged current $I=1$ partner,
measured in hadronic $\tau$ decay, by CVC. Recent
attempts to incorporate $\tau$ decay data into the evaluation
of $[ a_\mu ]_{had}^{LO}$ using the isospin-breaking (IB) corrected
version of this relation~\cite{dehz03} have encountered a 
consistency problem in the dominant 
($\pi\pi$ final state) contribution to $[ a_\mu ]_{had}^{LO}$.
Specifically, even after accounting for known sources of IB~\cite{cen}, 
the $\pi\pi$ component of the resulting
$\tau$-based $I=1$ vector spectral function is incompatible
with that obtained from the high-precision CMD2 
$e^+e^-\rightarrow\pi^+\pi^-$ data~\cite{cmd203}, with surprisingly
large ($\sim 10\%$) discrepancies remaining
in the region of $\pi \pi$ invariant masses from $\sim 0.85$
to $\sim 1$ GeV~\cite{dehz03}. SM predictions for $a_\mu$ based solely on
electroproduction (EM) data indicate a $\sim 2.5\sigma$ deviation from the 
experimental result, while those based on the IB-corrected hadronic
$\tau$ decay data indicate compatibility at the $\sim 1\sigma$
level~\cite{dehz03,otherSMamu}. The discrepancy thus significantly impacts
the question of whether or not the current experimental determination
of $a_\mu$~\cite{bnlamu} shows evidence for beyond-the-SM contributions. 

To resolve the above discrepancy, we have investigated sum rule
constraints on the EM and $\tau$-decay-based data sets, working with sum rules 
of the form $\int_{s_{th}}^{s_0}ds\, w(s)\, \rho (s)\, 
=\, {\frac{-1}{2\pi i}}\, \int_{\vert s\vert =s_0}ds\, w(s)\, \Pi (s)$,
where $s_{th}$ is the relevant threshold, 
$\Pi (s)$ is the correlator with spectral function
$\rho (s)$, $w(s)$ is a function analytic in the region
of the contour, and the OPE representation of $\Pi (s)$
is to be used on the RHS. At intermediate scales, 
weights satisfying $w(s=s_0)=0$ must be employed, in order
to suppress duality violating contributions from the vicinity of
the timelike point~\cite{pinchedfesrs}. 
This is most conveniently done by using the variable $y=s/s_0$
and weights $w(y)$ satisfying $w(1)=0$. Integrated OPE contributions
of dimension $D$ then scale as $1/s_0^{(D-2)/2}$,
allowing reliable self-consistency checks for the absence
of neglected higher $D$ contributions. 

In the current study, to be specific, we employ the
ALEPH data and covariance matrix~\cite{ALEPH97} 
(with updated normalization) for the hadronic $\tau$
decay version of the isovector vector spectral function.
For the various exclusive components of the EM spectral function 
and their errors, we follow the assessments of Refs.~\cite{dehz03,whalley03}.
More recent high-precision data is incorporated as discussed in 
Ref.~\cite{kmamu}. In the region 
of the EM-$\tau$ discrepancy, we focus on the CMD2 $\pi\pi$ data. 
The weights, $w(y)$, and scales, $s_0$, used in this study 
have been chosen in such a way that (i) the OPE is essentially 
entirely dominated by its $D=0$ contribution, (ii) the convergence of the
integrated $D=0$ series, order-by-order in $\alpha_s$, is excellent,
and (iii) poorly known $D=6$ contributions are absent~\cite{kmamu}. 
The OPE sides of the various sum rules are then determined,
up to very small non-perturbative corrections, by
the single input parameter, $\alpha_s(M_Z)$, for which
we employ an average of high-scale determinations
independent of the EM and $\tau$ data being tested~\cite{kmamu}.
The values relevant to lower scales are obtained by standard
four-loop running and matching~\cite{chet4looprun}.

For illustration, we present below results for three weight cases, $w(y)=1-y$,
$w_3(y)$, and $w_6(y)$, where $w_N(y)=1-\left({\frac{N-1}{N}}\right)y
+{\frac{y^N}{N}}$. The first weight has a zero of order $1$ at $y=1$,
the remaining weights zeros of order $2$. The weights are all
non-negative in the integration region, simplifying the interpretation
of the sum rule tests. A discussion of other advantages
of these weights may be found in Ref.~\cite{kmamu}.

\section{Results and Discussion}

We find that the $\tau$-decay-based
spectral integrals are compatible with the high-scale OPE input
for a wide range of $s_0$, but that the
EM data yields spectral integrals consistently lower than
those predicted by the OPE~\cite{kmamu}. The $s_0$ dependence of the
OPE and corresponding spectral integrals is also in good
agreement for the $\tau$ case, but in significant disagreement for the 
EM case~\cite{kmamu}. Figures may be found in Ref.~\cite{kmamu}. 

Regarding the normalization of the EM spectral integrals: 
to quantify the normalization discrepancy, we have used 
the spectral integral values at the highest accessible common
scale for the EM and $\tau$ cases ($s_0=m_\tau^2$) to obtain
an effective value for $\alpha_s(M_Z)$. This is to be compared 
to the independent high-scale average, $\alpha_s(M_Z)=0.1200\pm 0.0020$. 
The results are shown in Table 1. 
While the values implied by the EM data are low by only
$\sim 2\sigma$, typical fluctuations in the data
which would bring them into better agreement with OPE
expectations would also serve to increase 
$\left[ a_\mu\right]_{had}^{LO}$. The $\tau$ data, in contrast,
yields $\alpha_s(M_Z)$ values in good agreement with
the high-scale average. Somewhat lower values
of $\alpha_s(M_Z)$ are also easily accommodated since the
overall normalization uncertainty for the $\tau$ data
corresponds to an uncertainty of $\pm 0.0010$ in the
fitted $\alpha_s(M_Z)$.

\begin{table}[h]
\tbl{Values of $\alpha_s(M_Z)$ obtained by
fitting to the $s_0=m_\tau^2$ experimental EM and $\tau$ spectral integrals
with central values for the $D=2,4$ OPE input.}
{\begin{tabular}{@{}ccc@{}}\toprule
Weight&\qquad\qquad\qquad\qquad\qquad\qquad EM or $\tau$\ \ \ 
\qquad\qquad\qquad\qquad\qquad\qquad&$\alpha_s(M_Z)$\\
\hline
$1-y$&EM&$0.1138^{+0.0030}_{-0.0035}$\\
$w_3$&EM&$0.1152^{+0.0019}_{-0.0021}$\\
$w_6$&EM&$0.1150^{+0.0022}_{-0.0026}$\\
\hline
$1-y$&$\tau$&$0.1218^{+0.0027}_{-0.0032}$\\
$w_3$&$\tau$&$0.1195^{+0.0018}_{-0.0021}$\\
$w_6$&$\tau$&$0.1201^{+0.0020}_{-0.0022}$\\
\botrule
\end{tabular}}
\end{table}

The problem of the $s_0$ dependence of
the EM spectral integrals is similarly quantified in Table 2,
for $w(y)=1-y,\ w_6(y)$. $S_{exp}$ and $S_{OPE}$
are the slopes with respect to $s_0$
of the spectral and OPE integrals, respectively.
OPE entries labelled ``indep'' correspond to the high-scale average 
$\alpha_s(M_Z)$ input above, those labelled ``fit'' to
the fitted values given in Table 1.
The uncertainty in $S_{OPE}$ is seen
to be very small. The very weak dependence on $\alpha_s$ means that
even the change from ``indep'' to ``fit'' input on the OPE 
side has only marginal impact, the OPE versus
spectral integral slope discrepancy
being reduced from $2.6$ to $2.3\, \sigma$ for $w(y)=1-y$ 
and $2.5$ to $2.2\, \sigma$ for $w(y)=w_6(y)$.
The source of the slope problem thus lies entirely on the data side.
The problem can be cured only by a change in the shape of the EM
spectral distribution, such as would occur if
the IB-corrected $\tau$ data, rather than the EM data, represented
the correct version of $\rho_{EM}^{I=1}(s)$.

\begin{table}[ht]
\tbl{Slopes wrt $s_0$ of the EM OPE and spectral integrals}
{\begin{tabular}{lccc}
\hline
Weight&$S_{exp}$&\qquad$\alpha_s(M_Z)$\qquad&$S_{OPE}$\\
\hline
$1-y$&$.00872\pm .00026$&indep&$.00943\pm .00008$\\
&&fit&$.00934\pm .00008$\\
\hline
$w_6$&$.00762\pm .00017$&indep&$.00811\pm .00009$\\
&&fit&$.00805\pm .00009$\\
\hline
\end{tabular}}
\end{table}

We conclude that either (i) non-one-photon
physics effects, as yet unidentified, are contaminating the EM
data or (ii) there remain experimental problems with the 
(pre-2005) EM data. The latter conclusion is favored by the just-released SND
$e^+e^-\rightarrow\pi^+\pi^-$ results~\cite{snd05}, which
are compatible with the IB-corrected hadronic $\tau$ decay data. In
either case it follows that determinations of
$[a_\mu ]_{had}^{LO}$ incorporating $\tau$ decay data are
favored over those based solely on EM cross-sections
and hence that the SM prediction for $a_\mu$ is 
in good agreement with the current experimental result.

A final comment concerns the IB correction associated 
with $\rho$-$\omega$ ``mixing''. The current
standard evaluation~\cite{cen} is based on a chirally-constrained
model (the ``GP/CEN model'') fitted to a since-corrected form of the
CMD2 data. Because of strong cancellations, the model dependence of the
integrated $\left[ a_\mu\right]_{had}^{LO}$ interference contribution
turns out to be significantly larger than any one model's
fitting-induced uncertainty~\cite{kmcw05}. Table 3 shows 
the results of fits to the CMD2 data
for the Gounaris-Sakurai (GS), hidden local symmetry (HLS),
and Kuhn-Santamaria (KS) models, as well as
two versions of the GP/CEN model~\cite{cen}, one a
refitting to the current version of the CMD2 data (GP/CEN$^\dagger$), 
the other a refitting which incorporates an additional phase, 
such as arises when direct IB $\omega\rightarrow\pi\pi$ decay 
contributions are taken into account (GP/CEN$^*$). Smaller
values of $\left[\delta (a_\mu )\right]_{\rho -\omega}$ also serve to
reduce the difference between the EM and $\tau$-based
determinations of $\left[ a_\mu \right]_{had}^{LO}$.

\begin{table}[ht]
\tbl{Model dependence of the integrated $\rho$-$\omega$ interference
IB correction}
{\begin{tabular}{lcc}
\toprule
Model&\qquad\qquad\qquad\qquad$\chi^2/dof$\qquad\qquad\qquad\qquad\qquad&
$\left[\delta (a_\mu )\right]_{\rho -\omega}
\times 10^{10}$ \\
\hline
GS&35.9/38&$2.0\pm 0.5$\\
HLS&36.6/38&$4.0\pm 0.6$\\
KS&37.1/38&$3.8\pm 0.6$\\
GP/CEN$^*$&40.6/39&$2.0\pm 0.5$\\
GP/CEN$^\dagger$&61.5/40&$3.7\pm 0.7$\\
\botrule\end{tabular}}
\end{table}
%$^\dagger$ Modified version of the GP/CEN model with additional
%phase
%\vskip .05in
%$^*$ Original version of the GP/CEN model (with no additional phase)
%re-fitted to account for the 2003 correction to the CMD2 data
%\end{center}

\section*{Acknowledgements}

The hospitality of the CSSM, University of Adelaide,
and ongoing support of the Natural Sciences and Engineering
Council of Canada, are gratefully acknowledged.

%%%%%%%%%%%%%%%%%%%%%%%%%%%%%%%%%%%%%%%%%%%%%%%%%%%%%%%%%%%%
% Doing references:	                   	           %
%%%%%%%%%%%%%%%%%%%%%%%%%%%%%%%%%%%%%%%%%%%%%%%%%%%%%%%%%%%%

\end{document}